\documentclass{emulateapj}
\usepackage{natbib}
\usepackage{times}
\usepackage{graphicx}
\begin{document}

\shorttitle{Revealing Type Ia supernova physics}
\shortauthors{HORIUCHI and BEACOM}
\title{Revealing Type Ia supernova physics with cosmic rates and nuclear gamma rays}
\author{Shunsaku Horiuchi\altaffilmark{1,2} and John F. Beacom\altaffilmark{1,2,3}}

\altaffiltext{1}{Dept.\ of Physics, The Ohio State University, 191 W.\ Woodruff Ave., Columbus, OH 43210}
\altaffiltext{2}{Center for Cosmology and Astro-Particle Physics, The Ohio State University, 191 W.\ Woodruff Ave., Columbus, OH 43210}
\altaffiltext{3}{Dept.\ of Astronomy, The Ohio State University, 140 W.\ 18th Ave., Columbus, OH 43210}

\email{horiuchi@mps.ohio-state.edu}
\email{beacom@mps.ohio-state.edu}

\begin{abstract}
Type Ia supernovae (SNIa) remain mysterious despite their central importance in cosmology and their rapidly increasing discovery rate. The progenitors of SNIa can be probed by the delay time between progenitor birth and explosion as SNIa. The explosions and progenitors of SNIa can be probed by MeV nuclear gamma rays emitted in the decays of radioactive nickel and cobalt into iron. We compare the cosmic star formation and SNIa rates, finding that their different redshift evolution requires a large fraction of SNIa to have large delay times. A delay time distribution of the form $t^{-\alpha}$ with $\alpha = 1.0 \pm 0.3$ provides a good fit, implying $50\%$ of SNIa explode more than $\sim 1$ Gyr after progenitor birth. The extrapolation of the cosmic SNIa rate to $z = 0$ agrees with the rate we deduce from catalogs of local SNIa. We investigate prospects for gamma-ray telescopes to exploit the facts that escaping gamma rays directly reveal the power source of SNIa and uniquely provide tomography of the expanding ejecta.  We find large improvements relative to earlier studies by Gehrels et al.~in 1987 and Timmes \& Woosley in 1997 due to larger and more certain SNIa rates and advances in gamma-ray detectors. The proposed Advanced Compton Telescope, with a narrow-line sensitivity $\sim 60$ times better than that of current satellites, would, on an annual basis, detect up to $\sim 100$ SNIa ($3\sigma$) and provide revolutionary model discrimination for SNIa within 20 Mpc, with gamma-ray light curves measured with $\sim 10\sigma$ significance daily for $\sim 100$ days.  Even more modest improvements in detector sensitivity would open a new and invaluable astronomy with frequent SNIa gamma-ray detections.
\end{abstract}

\keywords{
Gamma rays: diffuse background ---
Gamma rays: stars ---
Nuclear reactions, nucleosynthesis, abundances ---
Stars: statistics ---
Supernovae: general
}

\section{Introduction}

Type Ia supernovae (SNIa) are deeply connected with many important frontiers of astrophysics and cosmology. They occur in all galaxy types and are major contributors to galactic chemical evolution, in particular of iron \citep[e.g.,][]{1986A&A...154..279M}. They are very bright and, as high redshift distance indicators, play a critical role in establishing the modern cosmology paradigm \citep{1998AJ....116.1009R,1999ApJ...517..565P,2001ApJ...553...47F}. 

However, there are major uncertainties regarding the nature of the SNIa progenitors and explosions. While it is established that most SNIa result from the thermonuclear explosion of carbon-oxygen white dwarfs (WD) near the Chandrasekhar mass, the mechanism of mass gain remains debated. In the single-degenerate (SD) scenario the WD accretes mass from a companion star \citep{1973ApJ...186.1007W}, while in the double-degenerate (DD) scenario the WD merges with another WD \citep{1984ApJS...54..335I}. In addition, although the main products are known, the basic mechanism of nuclear burning remains under debate. In deflagration the ignited flame propagates subsonically, while in detonation it propagates supersonically as a shock wave. There are combined models, as well as the possibility of detonation in the He layer of a sub-Chandrasekhar WD \citep[see, e.g.,][]{2000ARA&A..38..191H,2007ARep...51..291T}.

The difference between star formation and SNIa rates depends on what the SNIa progenitors are. The rate of SNIa following an episode of star formation depends on (i) the delay time, which describes the time required for a newly-formed SNIa progenitor to develop into a SNIa, and (ii) the production efficiency, which is the number of SNIa produced per $\rm M_\odot$ of star formed \citep[e.g.,][]{2005A&A...441.1055G}. The SD and DD scenarios involve different progenitors and have different predictions for these quantities. Although current predictions are comparable within uncertainties, improved information on the global delay-time distribution (DTD) and SNIa efficiency will discriminate between SNIa progenitor scenarios \citep{2000ApJ...528..108Y}. 

It is well known from nuclear fusion modeling and optical light curve observations that each SNIa yields as much as 0.5--0.7 ${\rm M_\odot}$ of radioactive $^{56}$Ni. The decay of $^{56}$Ni, via $^{56}$Co to stable $^{56}$Fe, provides the primary source of energy---in the form of nuclear gamma rays and energetic positrons that deposit much of their energy in the ejecta---that powers the SNIa optical display. Initially, the gamma-rays are trapped, but as the SNIa ejecta expands and the matter density drops they start to escape, while the positrons remain largely trapped until much later times. The decline rates of the optical light curve show the timescales set by the decay of $^{56}{\rm Ni}$ (half life of 6 days) and $^{56}{\rm Co}$ (half life of 77 days).

The detection of gamma rays that escape the ejecta is the key to resolving the central mysteries of SNIa \citep{1994ApJS...92..351G,1998ApJ...492..228H,2006NewAR..50..604B}. Since gamma rays are more penetrating and their opacities are much simpler than those for optical photons, they offer a more straightforward and direct probe of the inner mechanisms of SNIa \citep{1969ApJ...155...75C,2004ApJ...613.1101M,2008NewAR..52..377I}. The gamma ray flux allows the identification of radioactive material yield, and its time evolution allows tomography of the surrounding ejecta; both quantities can be compared with model predictions. In a few cases of fortuitously nearby SNIa, limits just above theoretical expectations have been set \citep{1990ApJ...362..235M,1994A&A...292..569L,1999HEAD....4.0803L}. No SNIa features have been seen in the cosmic gamma-ray background \citep[CGB; e.g.,][]{2005JCAP...04..017S}. Reliable knowledge of SNIa rates is essential to defining realistic prospects for gamma-ray detection, and new data make this possible .
 
In the first part of this paper, we investigate the cosmic SNIa rate and determine the delay of SNIa relative to their progenitor formation and the efficiency of forming SNIa (Section \ref{sec:progenitor}). We use a large selection of star formation indicators with recent updates (Section \ref{sec:rate}), as well as a comprehensive compilation of SNIa rate data (Section \ref{sec:ratedata}). We study the effects of delay and efficiency using data over a substantial redshift range, and we discuss what our results reveal about the SNIa progenitors, i.e., for SD and DD scenarios (Section \ref{sec:ratesynthesis}).

In the second part of this paper, we bring in gamma rays as a probe of SNIa explosions and progenitors (Section \ref{sec:gamma}). Our analysis of the cosmic SNIa rate plays a critical role in determining the gamma-ray detection prospects. We first review the gamma-ray emission for several benchmark SNIa models (Section \ref{sec:gammaemission}). We determine the SNIa contribution to the CGB using our updated cosmic SNIa rate inputs (Section \ref{sec:cgb}). We derive SNIa rates from supernova catalogs, and joining them with cosmic SNIa rates, we investigate the local SNIa rate with a higher accuracy than previously possible (Section \ref{sec:catalog}). We revisit the status of gamma-ray observations in light of our updated local SNIa rate (Section \ref{sec:localpast}). Finally, we give new results on the excellent gamma-ray detection prospects for nearby SNIa and the physics that this will probe (Section \ref{sec:local}). 

We close with a summary of our findings (Section \ref{sec:summary}). Throughout, we adopt the standard $\Lambda$CDM cosmology with $\Omega_m=0.3$, $\Omega_\Lambda=0.7$, and $H_0=70$ km s$^{-1}$ Mpc$^{-1}$.

\section{SNIa progenitors and cosmic SNIa rate}\label{sec:progenitor}

Comparing the cosmic star formation rate and the high redshift SNIa rate of the \emph{GOODS} fields suggests a DTD tightly distributed around 3--4 Gyr \citep{2004ApJ...613..200S,2008ApJ...681..462D,2010ApJ...713...32S}. However, small SNIa statistics, uncertain dust corrections, and the uncertain star formation rate at high redshift complicate studies of the DTD in this range \citep{2008MNRAS.388..829G}. We perform a new investigation of the SNIa rate by adopting both greatly updated star formation rate data (Section \ref{sec:rate}) and SNIa rate data (Section \ref{sec:ratedata}). In particular, we focus on the low redshift range, which provides a complementary test to previous studies. By comparing the calculation to observations, we determine the global DTD and SNIa efficiency, and discuss implications for SNIa progenitors (Section \ref{sec:ratesynthesis}). 

\subsection{SNIa rate calculation \label{sec:rate}}

The SNIa rate at a given time corresponds to the stellar birthrate at earlier epochs, convolved with the DTD, normalized by the efficiency of forming SNIa progenitor systems. The normalization is
\begin{equation}
\eta = \int {\rm A}_{\rm Ia}(M) \xi(M)\,dM,
\end{equation}
where $A_{\rm Ia}(M)$ is the SNIa formation efficiency, $\xi(M)$ is the initial mass function, and the integration is performed over the mass range for SNIa progenitors. Unless otherwise stated, we present all results for the Salpeter initial mass function; our results on the SNIa rate do not strongly depend on this choice because the SNIa progenitor mass range is sufficiently high. Assuming ${\rm A_{Ia}}$ and $\xi$ do not vary with space and time, $\eta$ is a constant representing the number of SNIa produced per $\rm M_\odot$ of stars formed, so that
\begin{equation} \label{SNIarate}
R_{\rm Ia} \left[ z(t_c) \right] = \eta \int_{t_{10}}^{t_c} \Phi(t_c-t'_c) \dot{\rho}_\star[z(t'_c)] \, dt'_c, 
\end{equation}
where $\Phi(t)$ is the DTD, its argument is the time difference between progenitor star formation and SNIa explosion, $ \dot{\rho}_{\star}(t_c) $ is the star formation rate, and the integration is performed over cosmic time from $t_{10}$, the age of the universe when the first stars were being formed, $z \approx 10$. $\Phi(t)$ is normalized such that its integral over the entire delay time range is unity. We do not consider the effect of metallicity on the rates \citep[e.g.,][]{2003A&A...406..259P,2008ApJ...673..999P,2009A&A...503..137B,2009MNRAS.395.2103M}.

\begin{figure}[t]
\centering\includegraphics[width=\linewidth,clip=true]{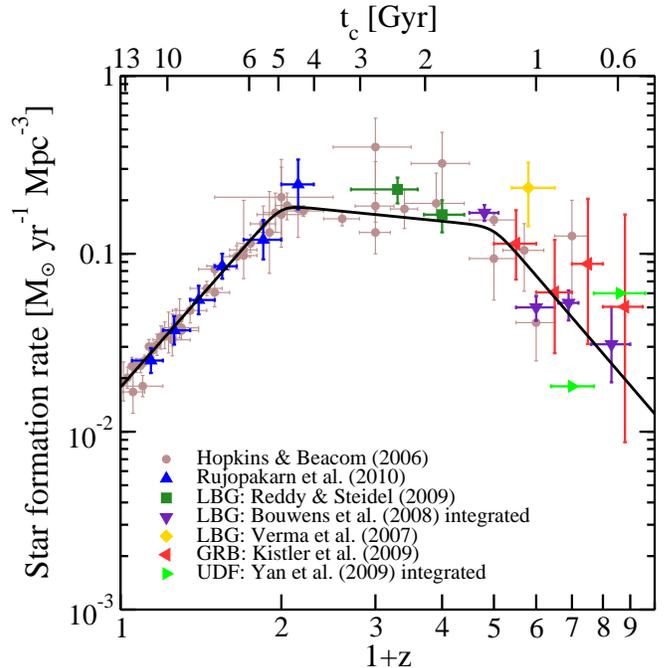}
\caption{Comoving cosmic star formation rate density as a function of $(1+z)$ on a logarithmic axis. Data as labeled (see text for details), agree to within $\pm 20$\% for $z \lesssim 1$. The uncertainty increases with redshift, but the overall shape is well constrained. The fitted curve is shown by a solid line \citep{2008ApJ...683L...5Y}. \label{fig:sfr}}
\end{figure}

In the limit that $\Phi(t)$ is large only for small $t$, we recover a prompt population of supernovae, i.e., similar to core-collapse supernovae, for which the cosmic rate evolution is the same as for the star formation rate. For SNIa, the observation that they occur in both young star-forming galaxies and old elliptical galaxies with little star formation demonstrates the need for a wide range of delay times. Recently, two populations of SNIa, a ``prompt'' population with delay times $\sim 0.1$ Gyr and a ``delayed'' population with delay times $\sim 10$ Gyr, have been reported \citep{2005A&A...433..807M,2005ApJ...629L..85S,2006MNRAS.370..773M,2006ApJ...648..868S,2010arXiv1002.3056M}. Whether this represents a truly bimodal distribution, or whether it is the result of two coarsely sampled bins of what is in reality a continuous DTD, remains uncertain. 

We adopt a power-law DTD, $\Phi(t) \propto t^{-\alpha}$. Power-laws have been reported from SNIa observations: for example, with $\alpha = 0.5 \pm 0.2$ from the SNLS data \citep{2008ApJ...683L..25P}, with $\alpha = 1.08^{+0.15}_{-0.11}$ from observations of transients in passively evolving galaxies \citep{2008PASJ...60.1327T}, with $\alpha \sim 1$ from local SNIa in the Lick Observatory Supernova Search (LOSS) \citep{2010arXiv1002.3056M}, and most recently $\alpha= 1.2 \pm 0.3$ from cluster SNIa \citep{2010arXiv1006.3576M}. In fact, the number of WDs following an instantaneous star formation burst is also a simple power-law in $t$: for an initial mass function $\xi(M) \propto M^{x}$ ($x=-2.35$ for Salpeter) and a power-law approximation to the evolutionary time-scale of 3--8 $\rm M_\odot$ stars $t_{\rm age} \propto M^y$ ($y \approx -2.6$), the index is $\alpha = (x-y+1)/y \approx 0.5$. However, this distribution extends only over the lifetimes of progenitors, 30--400 Myr, which is much smaller than the observed range of delay times. The extra time needed for mass transfer before triggering a SNIa would shift the delay time distribution towards longer delays, but the final DTD is largely dependent on the properties of the binary and its evolution.

We apply our DTD over times 0.1--13 Gyr and adopt $\alpha$ as a parameter. The lower time limit could be as short as the lifetime of a 8 $\rm M_\odot$ star, $\sim$ 0.03~Gyr, but recent studies of the spatial distribution of SNIa in their host galaxies indicate a lower limit of 0.2~Gyr \citep{2009ApJ...707...74R}. We adopt 0.1 Gyr; at the precision of the current SNIa rate measurements, our results are not very sensitive to small delays. We fix our upper limit to the look-back time to $z=10$, which is sufficiently long to accommodate SNIa observed in the oldest elliptical galaxies. Allowing a larger value would not change the redshift evolution, and would simply scale the efficiency. 

The cosmic star formation rate density is shown in Figure~\ref{fig:sfr}. The compilation of  \cite{2006ApJ...651..142H}, which includes data from multiple star formation indicators, is shown together with recent data from mid-infrared-emitting galaxies \citep{2010ApJ...718.1171R}, Lyman break galaxies (LBG) \citep{2007MNRAS.377.1024V,2007ApJ...670..928B,2008ApJ...686..230B,2009ApJ...692..778R}, and gamma-ray burst (GRB) host galaxies \citep{2008ApJ...683L...5Y,2009ApJ...705L.104K,2010MNRAS.406.1944W,2010ApJ...711..495B}. Note that \cite{2007ApJ...670..928B,2008ApJ...686..230B} report results integrated down to a chosen luminosity limit; we show instead complete integrations (see \cite{2009ApJ...705L.104K} for details). We also show Hubble Ultra Deep field (UDF) measurements from \cite{2010RAA....10..867Y}, similarly integrated to low galaxy luminosities. New UDF results from \cite{2010arXiv1006.4360B} have just appeared. As these indicate a steeper rise at low galaxy luminosities, the integration for the full star formation rate would need to be carefully regulated, and we do not perform this here. The star formation rate has been checked by independently measured quantities, such as the extragalactic background light \citep{2009PhRvD..79h3013H}, stellar mass density \citep{2008MNRAS.385..687W}, and upper limits on the diffuse supernova neutrino background \citep{2010arXiv1004.3311B}. 

On the top axis we label cosmic time. It highlights that delay times of Gyrs will result in noticeable differences in the slope between the star formation and SNIa rate rates. The star formation rate rapidly increases by one order of magnitude between redshift 0 and 1, which has been very well measured by a variety of indicators. This serves as an important feature to be compared to the SNIa rate. At higher redshift, the star formation rate eventually declines, at first slowly after $z=1$, and then more rapidly after $z \sim 4$. The exact slope above $z \sim 4$ remains uncertain, although this does not affect our conclusions, as we discuss in Section \ref{sec:ratesynthesis}.

\subsection{SNIa rate measurements \label{sec:ratedata}}

Since the first measurement of SNIa at cosmological distances \citep{1996ApJ...473..356P}, many surveys have collected homogenous samples of CCD-discovered SNIa. These surveys periodically observe the same patch of sky, or the same sample of galaxies, to find transients. Cuts are applied to select the most confidently-identified supernovae, and corrections for dust and incompleteness are applied. We summarize the present SNIa rate measurements in Table~\ref{table:data}. 

The high redshift ($z\gtrsim 0.5$) rates are the hardest to measure. Dust corrections are non-trivial and uncertain, e.g., the dust corrected data of \cite{2004ApJ...613..189D,2008ApJ...681..462D} are a factor $\sim 2$ higher than their uncorrected data. Even this has been argued to be insufficient and the rates could be treated as lower limits \citep{2008MNRAS.388..829G}. On the other hand, SNIa surveys are likely contaminated by core-collapse supernovae, for which the rate increases from $\approx 10^{-4} \,{\rm  yr^{-1} \,  Mpc^{-3}}$ at $z=0$ to $\approx 10^{-3} \, {\rm yr^{-1} \, Mpc^{-3}}$ by $z=1$, i.e., $\sim 10$ times the SNIa rate by $z=1$. It is therefore important that only the most confidently-identified SNIa---those that are spectroscopically confirmed---are selected for rate measurements. Indeed, rates derived using photometric classification \citep{2006ApJ...637..427B,2007MNRAS.382.1169P,2008ApJ...673..981K} are generally higher than those based on a higher fraction of spectroscopically confirmed SNIa \citep{2004ApJ...613..189D,2008ApJ...681..462D}. This is usually reflected in the large systematic errors. Finally, the sample sizes rapidly become small at high redshift, where typically $N_{\rm Ia}=$ 2--3. These issues make it difficult to use the high redshift rate measurements \citep{2008PASJ...60..169O}, even though they would be very sensitive to delays due to the small cosmic time difference since the first stars were being formed.

At intermediate redshift ($0.1 \lesssim z \lesssim 0.5$), many of the measurements are based solely on spectroscopic confirmed SNIa \citep{2000A&A...362..419H,2002ApJ...577..120P,2003ApJ...594....1T,2003ApJ...599L..33M,2004A&A...423..881B,2006AJ....132.1126N,2008ApJ...682..262D}. Multi-band photometric identification is also now reliable: \cite{2010ApJ...713.1026D} show that in the SDSS-II Supernova Survey only $\sim$2\% of the photometric SNIa in $z<0.3$ may be misidentified. Furthermore, the effects of dust become less important than at high redshifts \citep{2008A&A...479...49B}. However, many surveys target a pre-selected sample of galaxies and, while a large sample is adopted, there would be a bias towards bright galaxies, and supernovae that explode in faint galaxies would be missed. Lastly, we note that sample sizes vary largely from survey to survey.

Finally, in the local ($D \lesssim 100$ Mpc) volume, the rate has been calculated from compilations of carefully selected SNIa \citep{1999A&A...351..459C,2009MNRAS.395.1409S}, as well as the local LOSS measurement \citep{2010arXiv1006.4611L,2010arXiv1006.4612L,2010arXiv1006.4613L}. Note that the rate calculated from \cite{2009MNRAS.395.1409S} probably reflects a local enhancement (Section \ref{sec:catalog}).

\subsection{SNIa rate synthesis \label{sec:ratesynthesis}}

\begin{figure}[t]
\centering\includegraphics[width=\linewidth,clip=true]{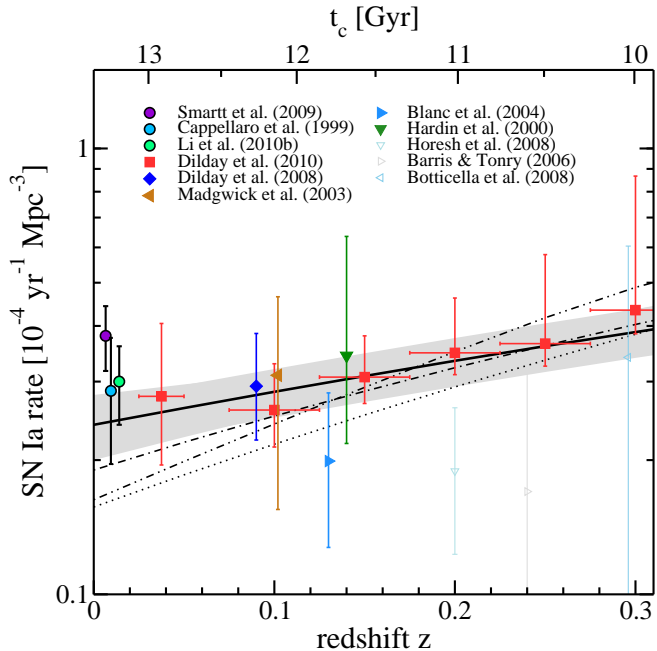}
\caption{Comoving SNIa rate as a function of redshift. Measurements are shown with statistical and systematic uncertainties combined in quadrature. The filled points are more reliable than the empty points (see text). The local \cite{2009MNRAS.395.1409S} point is affected by a local enhancement (see Section \ref{sec:catalog}). The best-fit $\alpha=1.0$ power-law DTD (solid) is shown, with a shaded band reflecting the uncertainty. Other DTD shown are as labeled in Figure \ref{fig:cosmicrate}; they do not fit the data as well. \label{fig:localrate}}
\end{figure}

Compared to the star formation rate, the SNIa rate is further away from a consensus. Although the number of rate measurements has increased, the scatter is large, and at times, measurements in comparable redshifts are in direct disagreement. Bearing in mind the strengths and weaknesses of the data, we group the rate measurements into two categories:
\begin{itemize}
\item \emph{Filled points}: those based on dedicated surveys using only spectroscopically identified SNIa. We also include select surveys with less than 100\% spectroscopic identification, including the survey by SDSS which includes an order of magnitude more SNIa than any other survey \citep{2010ApJ...713.1026D}, and measurements by \cite{2004ApJ...613..189D,2008ApJ...681..462D} which are the most reliable measurements in the high redshift regime.
\item \emph{Empty points}: those based largely on photometrically identified SNIa.
\end{itemize}
We caution that even within each category, the sample size, observation schedule, limiting magnitude, and other conditions vary, and our description is simply an attempt to appreciate some of the important differences between measurements \citep{2008NewA...13..606B}. 

Now, we can make better sense of the calculated and measured SNIa rates despite their uncertainties. When we restrict ourselves to the filled points, we find that the SNIa rate increases by only a factor $\sim 2$--3 from redshift 0 to 0.8 or so. On the other hand, the cosmic star formation rate increases by a factor 10 from redshift 0 to 1. \emph{This implies that a large fraction of SNIa have cosmologically-large time delays}. 

\begin{figure}[t]
\centering\includegraphics[width=\linewidth,clip=true]{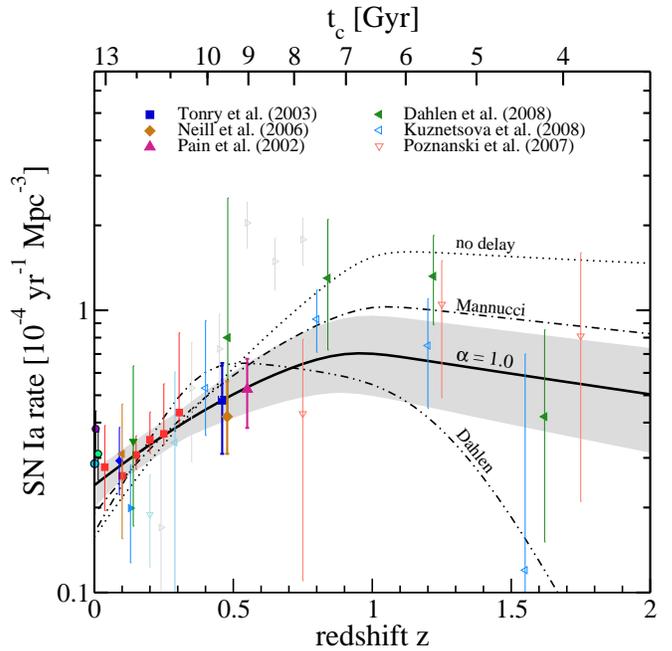}
\caption{Same as Figure \ref{fig:localrate}, but over the entire redshift range where data are available. In addition to the best-fit $\alpha=1.0$ power-law DTD (solid), we show the no delay (dotted), the two-component DTD of \cite{2006MNRAS.370..773M} (dot-dashed), and the 3.4 Gyr narrow DTD of \cite{2008ApJ...681..462D} (dot-dot-dashed), as labeled.
\label{fig:cosmicrate}}
\end{figure}

We show in Figures \ref{fig:localrate} and \ref{fig:cosmicrate} the calculated and measured SNIa rates. We fit the calculated cosmic SNIa rate form to the selection of rate data consisting of filled points and local measurements (excluding that by \cite{2009MNRAS.395.1409S}). For the power-law DTD we fit for the normalization at redshift zero, $R_{\rm Ia}(0)$, and the DTD exponent, $\alpha$. We find best-fit values $(R_{\rm Ia}(0) , \alpha)=(0.24,1.0)$ with $\chi^2=5.4$ for 17 degrees of freedom. The projections of the $\Delta \chi^2=2.30$ elliptical contour give $R_{\rm Ia}(0) = 0.24 \pm 0.04$ and $\alpha = 1.0 \pm 0.3$; these parameters are strongly anti-correlated. The uncertainty is shown as grey shading in the figures. The best-fit efficiency and uncertainty of making SNIa is $(5 \pm 1) \times 10^{-4} \, { \rm M_\odot^{-1}}$, or, assuming a SNIa progenitor mass range of 3--8 ${\rm M_\odot}$, a SNIa fraction (which we define as the number of SNIa divided by the number of 3--8 $\rm M_\odot$ stars formed; the fraction of stars in SNIa-producing binaries is twice this) of $2.4 \pm 0.5$\%. 

In addition to the power-law DTD we show the bimodal DTD of \cite{2006MNRAS.370..773M} (dot-dashed), the 3.4 Gyr narrow DTD of \cite{2004ApJ...613..200S,2008ApJ...681..462D,2010ApJ...713...32S} (dot-dot-dashed), and the no-delay case (dotted), for comparison. These DTDs do not fit the data as well, with $\chi^2$ values of 10, 17, and 21, respectively. All rise too fast compared to the reliable data (filled points). This also means they underpredict the $z=0$ rate. The different results in \cite{2004ApJ...613..200S,2008ApJ...681..462D,2010ApJ...713...32S} are due to the choice of data. The narrow DTD is driven by a declining SNIa rate at $z \sim 1.4$ (see green left-pointing triangles in Figure \ref{fig:cosmicrate}), while our result is driven by the slow rise at low redshift.

At present, the DTD does not clearly identify the SNIa progenitor scenario, because both the SD and DD scenarios can lead to power-law DTDs with $\alpha \approx 1$. In the DD scenario, the delay is approximately dictated by the time taken by a binary to merge by angular momentum loss, which from general relativity scales as the fourth power of the binary separation. For a logarithmically-flat distribution of binary separations, one obtains $\alpha \sim 1$. Although binary evolution synthesis studies of the SD scenario typically show that the DTD peaks at characteristic scales related to stable mass transfer \citep[e.g.,][]{2000ApJ...528..108Y,2005ApJ...629..915B,2009MNRAS.395.2103M,2010ApJ...710.1310M}, power-law DTDs have also been predicted: for example, \cite{2008ApJ...683L.127H} report power-law DTDs with $\alpha \approx 1$. Both progenitor scenarios are acceptable fits, but as predictions of SNIa progenitor scenarios improve and SNIa rate data accumulates, better testing would become possible. We have assumed for simplicity a power-law DTD; in order to generally test the progenitor scenario, a generic form for the DTD should be statistically tested \citep[e.g.,][]{2010ApJ...713...32S}. 

The choice of the initial mass function does not affect the shape of the SNIa rate, but changes the number of SNIa progenitor stars. This introduces a small tens of percent difference in the efficiency. For a modern Baldry-Glazebrook initial mass function with a low-mass suppression \citep{2003ApJ...593..258B}, the SNIa fraction required is $2.9\pm0.6$\%. Since the comparison of cosmic rates is not sensitive to small delays, there could be more short-lived progenitors than indicated by our simple DTD form. Uncertainty in the cosmic star formation rate beyond $z \sim 4$ affects the low redshift SNIa slope, albeit weakly, through the long tail of the DTD. There is a degeneracy in the high-redshift star formation slope and the value of $\alpha$, where a higher star formation rate can be compensated by a larger $\alpha$. However, at the precision of the current SNIa rate measurements, this uncertainty does not affect our conclusions. 

Our SNIa efficiency is comparable to those from recent studies of volumetric SNIa rates \citep{2008NewA...13..606B}, and somewhat lower than those derived by other methods \citep{2010MNRAS.tmp..906G,2010MNRAS.tmp..968M,2010arXiv1002.3056M}. Note that comparisons must be done with a common assumption for the initial mass function. As described above, a more modern initial mass function would increase our efficiency, although a difference remains. The true value of the efficiency remains to be determined; see \cite{2008MNRAS.384..267M} for a detailed discussion and implications of the SNIa efficiency. For example, our SNIa fraction contrasts with the $\sim 0.2$\% advocated in some DD scenarios, where conservative conditions on the mass ratios are applied \citep{2009ApJ...699.2026R}. The higher required efficiency that we found suggests the need to relax some assumptions, or to include contributions from SNIa of lower mass progenitors. We should note here that our definition for efficiency implicitly assumes binaries that would become a SNIa within $\approx 13$ Gyr.

It is revealing to discuss the SNIa rate jointly with the core-collapse supernova rate. The ratio of SNIa to core-collapse supernova (Ia/II) at $z=0$ is $\approx 0.2$--0.3. Here we use the $z=0$ cosmic SNIa rate instead of the local 10 Mpc data, as the latter rate is $\sim 0.1$ per year; the recent decade, for example, had no SNIa \citep{2008arXiv0810.1959K}. Since SNIa are delayed, the SNIa rate should increase less strongly than the star formation rate and the core-collapse rate from $z=0$ to $z=1$. The Ia/II ratio should therefore decrease with redshift in this range. However, the data do not clearly show such a trend, being consistent with no evolution (\cite{2009PhRvD..79h3013H}; this is true even with updates on the core-collapse rate, e.g., \cite{2009A&A...499..653B}). A likely possibility is that fainter core-collapse supernovae are being missed compared to the brighter SNIa \citep[see][]{2009PhRvD..79h3013H}. The importance of the missing core-collapse supernovae spans topics such as the formation of black holes in massive stellar collapse, supernova neutrinos, and metal enrichment.

\section{SNIa explosions and gamma-rays}\label{sec:gamma}

Gamma rays reveal information of the SNIa interior that can be effectively used to study SNIa physics. We first review the gamma-ray emission per SNIa, focusing on a range of SNIa models that accommodate normal, superluminous, and subluminous SNIa (Section \ref{sec:gammaemission}). Using the cosmic SNIa rate analyzed in Section \ref{sec:progenitor}, we discuss prospects for using the CGB to detect SNIa gamma rays (Section \ref{sec:cgb}). We then discuss the local SNIa rate with higher precision than previously possible, by combining our cosmic SNIa rate analysis with SNIa rates from supernova catalogs of nearby supernovae (Section \ref{sec:catalog}). After reviewing the current status of gamma-ray observations (Section \ref{sec:localpast}), we give new results detailing the prospects for SNIa gamma-ray detection and studying SNIa physics.

\subsection{Gamma-ray yield per SNIa \label{sec:gammaemission}}

\begin{table*}
\begin{center}
\caption{Gamma-ray line emission from SNIa \label{table:gamma}}
\begin{tabular}{llccccc}
\tableline\tableline
& & & & \multicolumn{3}{c}{Line flux over $10^6$ s [$10^{47} $ s$^{-1}$]} \\
\multicolumn{2}{c}{SNIa model} & Ref. & $M_{\rm Ni}$ [${\rm M_\odot}$] & 812 keV & 847 keV & 1238 keV \\
\tableline
Normal: delayed-detonation		& \emph{DD202C}	& 1	& 0.72	& $0.8$  	&  $5.2$   &  $3.9$  \\
Normal: deflagration				& \emph{W7} 		& 2 	& 0.58	& $0.5$  	&  $4.3$   &  $3.2$  \\
Normal: He-detonation			& \emph{HED8}	& 3	& 0.51	&   1.6	&   4.4	&	\\
\tableline
Superluminous: late-detonation	& \emph{W7DT} 	& 4	& 0.76 	&  2.2	&	5.9	&	\\
Superluminous: He-detonation		& \emph{HECD}	& 5	& 0.72 	&  2.5	&	5.5	&	\\
\tableline
Subluminous: He-detonation		& \emph{HED6}	& 3	& 0.26 	& 0.6		&	2.2	&	\\
Subluminous: pulsed delayed-detonation	& \emph{PDD54}	& 6	& 0.14 	& 0.05	&	1.2	&	\\
\tableline
\end{tabular}
\tablerefs{(1) \cite{1998ApJ...492..228H}; (2) \cite{1984ApJ...286..644N}; (3) \cite{1996ApJ...457..500H}; (4) \cite{1992ApJ...393L..55Y}; (5) \cite{1997thsu.conf..515K}; (6) \cite{1995ApJ...444..831H}.}
\end{center}
\end{table*}

In all SNIa models, the decay chain $^{56}{\rm Ni} \to \, ^{56}{\rm Co} \to \, ^{56}{\rm Fe}$ provides the primary source of energy that powers the SNIa optical display. The $^{56}$Ni decays by electron capture and the daughter $^{56}{\rm Co}$ emits gamma rays by the nuclear de-excitation process
\begin{eqnarray}\label{eq:Ni}
^{56}{\rm Ni} + e^- &\to& ^{56}{\rm Co}^* + \nu_e  \\ \nonumber
^{56}{\rm Co}^* &\to& ^{56}{\rm Co} + \gamma \\ \nonumber
E_\gamma &=& 158 \, {\rm keV} (99\%), \, 812 \, {\rm keV} (86\%),
\end{eqnarray}
where percentages express photons per decay and the sum can be larger than 100\%. Only the dominant lines are noted here. The daughter $^{56}$Co decays by electron capture (81\%) as well as positron emission (19\%), and the daughter $^{56}{\rm Fe}$ de-excites by emitting gamma rays
\begin{eqnarray}\label{eq:Co}
^{56}{\rm Co} + e^- &\to& ^{56}{\rm Fe}^* + \nu_e  \\ \nonumber
^{56}{\rm Co} &\to& ^{56}{\rm Fe}^* + e^+ + \nu_e  \\ \nonumber
^{56}{\rm Fe}^* &\to& ^{56}{\rm Co} + \gamma \\ \nonumber
E_\gamma &=& 847 \, {\rm keV} (100\%), \, 1238 \, {\rm keV} (67\%),
\end{eqnarray}
where the percentages include the effects of the 19\% branching ratio for positron production by beta decay. 

Initially, most of the gamma rays Compton scatter on the SNIa ejecta and deposit their energy. As the ejecta expands and the matter density decreases, more gamma rays escape. The observed 847 keV line luminosity is
\begin{equation} \label{eq:emission}
S_\gamma (t) = p_{\rm esc} (t) \left[ \frac{M_{\rm Ni}N_A}{56} \right] \left[ \frac{e^{-t/\tau_{\rm Co}} - e^{-t/\tau_{\rm Ni}} }{\tau_{\rm Co} - \tau_{\rm Ni}} \right]
\, {\rm \gamma \, s^{-1} },
\end{equation}
where $p_{\rm esc}$ is the model-dependent probability of escape through the SNIa ejecta, the quantity in the first square brackets defines the model-dependent nucleosynthesis yield, and the second square brackets reflect the nuclear decay rate. Here, $M_{\rm Ni}$ is the nickel mass, $N_A$ is the Avogadro number, and $\tau_{\rm Ni}= t_{1/2} / {\rm ln}(2)$ where $t_{1/2}$ is the half life:  $t_{1/2} = 6.1$~days for $^{56}$Ni and 77.2 days for $^{56}$Co. \emph{The distribution of $M_{\rm Ni}$ (over many SNIa) and especially the time-evolution of $p_{\rm esc}$ (per SNIa) can be used to probe SNIa physics.}

In deflagration, nuclear burning ignites near the center and burning moves subsonically across the progenitor. In pure detonation, the flame front propagates supersonically as a shock front, but we do not consider this further since the resulting elemental abundances disagree with data \citep{1971ApJ...165...87A}. In delayed-detonation, an initial deflagration becomes a detonation at some critical density that is an unknown parameter. All these models are usually assumed to be initiated from a WD near the Chandrasekhar mass, accreting mass from a non-degenerate star. In He-detonation, burning commences in the degenerate helium layer near the surface of a sub-Chandrasekhar mass WD. In DD scenario models, ignition occurs at low densities as mass from the disrupted binary accretes. It is expected that a large material envelope covers the burning sites, so that the escape probability for gamma rays is lower than in SD scenarios, thus probing the progenitors. 

\cite{2004ApJ...613.1101M} compared the gamma-ray emission from seven transport codes for a selection of SNIa models. They conclude that differences due to transport codes are 10--20\%, much less than the differences that result from models. The continuum emission is more model dependent than the line emission since it depends on multiple scatterings and the time-integrated continuum differs by up to a factor $\sim$ 2 between codes. 

From \cite{2004ApJ...613.1101M} we adopt a selection of SNIa models representative of normal, superluminous, and subluminous SNIa, and spanning the deflagration, delayed-detonation, and He-detonation models. The average peak line emission, over a $10^6$ s period, are summarized in Table~\ref{table:gamma}. The light curves for super-luminous SNIa (solid) and normal SNIa (dashed) are shown in Figure \ref{fig:lightcurve}. We comment on DD SNIa models in Section \ref{sec:discussion}. We adopt two time-integrated gamma-ray spectra: a deflagration model \emph{W7} of \cite{1984ApJ...286..644N}, which yields which yields $0.58 {\rm M_\odot}$ of $^{56}$Ni, and a delayed-detonation model \emph{5p0z22.23} of \cite{2002ApJ...568..791H}, which yields $0.56 {\rm M_\odot}$ of $^{56}$Ni. Both models are representative of normal SNIa, the most common kind. 

\subsection{Contribution to the cosmic gamma-ray background \label{sec:cgb}}

\begin{figure}[b]
\centering\includegraphics[width=\linewidth,clip=true]{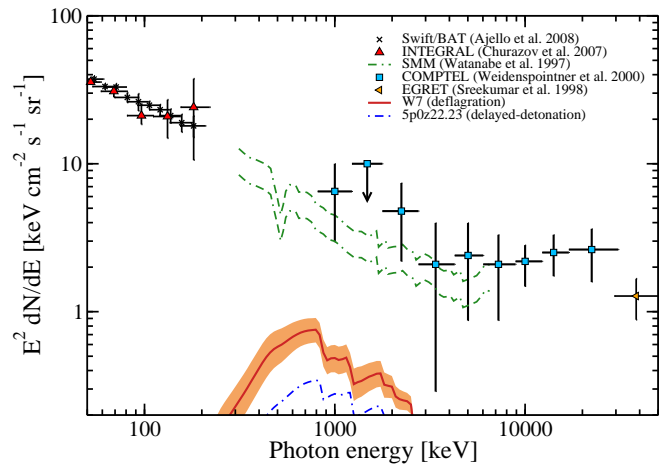}
\caption{The SNIa contribution to the cosmic gamma-ray background. The contribution from the deflagration model \emph{W7} (solid) and the delayed-detonation model \emph{5p0z22.23} (dot-dashed) are shown. For \emph{W7} we show the range owing to the uncertainty range of the SNIa rate (shaded); a similar range applies for \emph{5p0z22.23} but is not shown. Data are as labeled.
\label{fig:cgb}}
\end{figure}

The SNIa contribution to the CGB depends on the cosmic SNIa rate and the time-integrated gamma-ray number spectrum per SNIa, $f(E)$, as
\begin{equation}\label{eq:cgb}
E^2 \frac{{\rm d}N}{{\rm d}E}  =\frac{c}{4 \pi} \int^{z_{\rm max}}_0 R_{\rm Ia}(z) \frac{E^{\prime 2} }{(1+z)} f(E^\prime) \left| \frac{{\rm d}t}{{\rm d}z} \right| {\rm d}z,
\end{equation}
where $E$ is the measured photon energy, and $|{\rm d}z/{\rm d}t|=H_0(1+z) [\Omega_m(1+z)^3+\Omega_\Lambda]^{1/2}$. The left-hand side is equivalent to $\nu I_\nu$, and the redshift factor in Eq.~(\ref{eq:cgb}) comes from the energy scaling. 

In Figure~\ref{fig:cgb} we show the resulting SNIa contribution to the CGB, for the deflagration model \emph{W7} (solid) and the delayed-detonation model \emph{5p0z22.23} (dot-dashed). For \emph{W7}, the shading show the uncertainty due to SNIa rate. We see that the SNIa contributions are at least a factor $\sim 5$ smaller than current CGB measurements, and up to $\sim$20 depending on the SNIa rate and SNIa model. Although the spectrum per SNIa has line features, these are washed out due to redshift. Our results are comparable to those of previous studies. Although early studies showed large contributions from SNIa \citep{1969ApJ...158L..43C,1975ApJ...198..241C,1993ApJ...403...32T,1996MNRAS.281L...9Z,1999ApJ...516..285W,2001ApJ...549..483R}, later studies report contributions to be $\sim$10 \citep{2005JCAP...04..017S} and $\gtrsim$10 \citep{2005PhRvD..71l1301A} less than the measured intensity.

If the dominant fraction of the CGB could be subtracted by future detectors, the SNIa contribution could be detected. The feasibility depends on the true nature of the currently measured CGB. Proposed sources include various populations of AGNs \citep[e.g.,][]{2009ApJ...699..603A}, hot coronae of AGNs \citep{2008ApJ...672L...5I}, and exotic dark matter models \citep{2005PhRvD..72f1301A,2007PhRvL..99s1301C,2008JCAP...01..022L}. The measured CGB may also be dominated by detector backgrounds. The true nature is unknown, and remains an important quest for future experiments to elucidate. Angular-correlation techniques may help differentiate the various possibilities \citep{2004ApJ...614...37Z}. 

\subsection{Local SNIa rate measurements \label{sec:catalog}}

\begin{figure}[t]
\centering\includegraphics[width=\linewidth,clip=true]{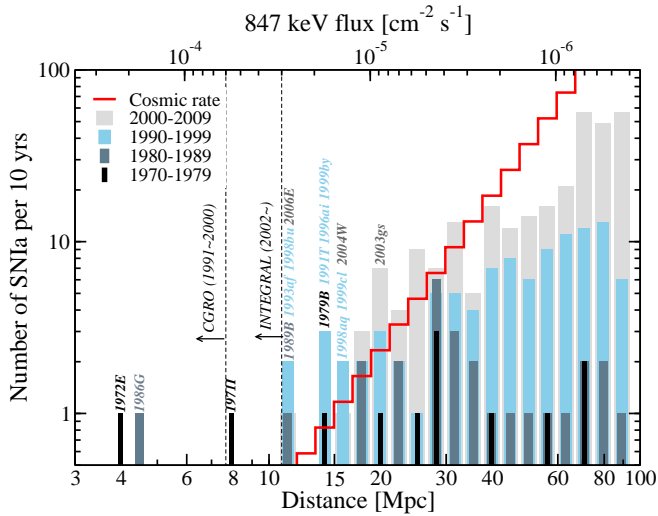}
\caption{Number of SNIa in the SAI catalog in 10 year bins, plotted differentially in distance, with selected SNIa as labeled. The solid step shows the cosmic SNIa rate extrapolated to the local volume, revealing the incompleteness of the catalog at large distances. On the top x-axis, the distance is converted to a 847 keV line flux, assuming $4.3 \times 10^{47} $ s$^{-1}$ at peak. The dashed lines represent $3 \sigma$ detector line sensitivities ($10^6$ s), labeled with years of operation; the sensitivity is dependent on exposure (see text). 
\label{fig:catalog}}
\end{figure}

Knowing the local SNIa rate is a critical prerequisite for assessing the prospects for detecting gamma rays from individual SNIa. Many studies have discussed the SNIa rate, most notably in the 1980s by \cite{1987ApJ...322..215G} and in the 1990s by \cite{1997ApJ...489..160T}, which provided much needed guidance on gamma-ray detection prospects. Now we have the advantage of systematic supernova surveys and greatly improved SNIa statistics. In addition to supernova catalogs, we use our cosmic SNIa rate to arrive at a consistent picture for the local SNIa rate.

More than 5000 supernovae discovered up to the end of 2009 are listed in the Sternberg Astronomical Institute Supernova Catalog \citep[SAI; ][]{2007HiA....14..316B}, the Asiago Supernova Catalog, and the catalog maintained by the Central Bureau for Astronomical Telegrams. In some cases, the catalogs disagree on details, but these discrepancies are increasingly rare for more recent supernovae. In the presence of a disagreement, we chose the classification in SAI; only small quantitative, and no qualitative, differences appear if preference is given to the other catalogs. We have used the reported recession velocities in the SAI catalog for their distances, with cross checks for the nearest SNIa with catalogs of galaxies \citep{2004AJ....127.2031K}.

\begin{figure}[t]
\centering\includegraphics[width=\linewidth,clip=true]{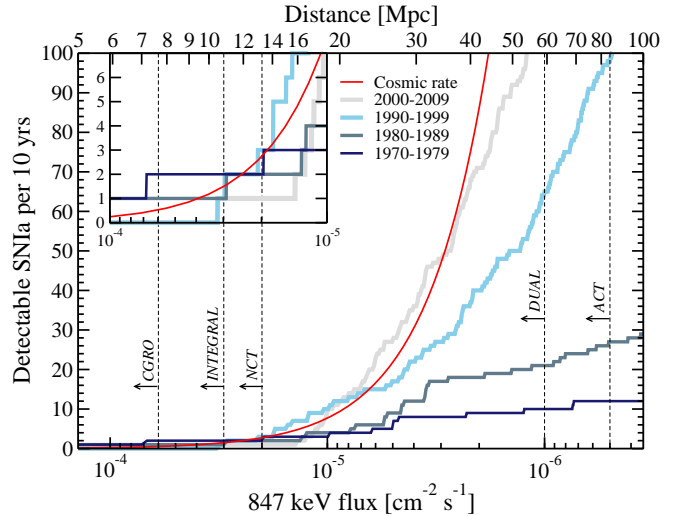}
\caption{The total (i.e. cumulative) number of gamma-ray detectable SNIa rate per 10 years, as a function of 847 keV line flux (bottom axis), or distance (top axis; assuming 847 keV peak emission of $4.3 \times 10^{47} $ s$^{-1}$ per SNIa). The extrapolation of the cosmic SNIa rate, as well as the rates derived from supernova catalogs, are shown and labeled accordingly. The dashed lines are $3 \sigma$ detector line sensitivities ($10^6$ s) as labeled. The inset shows part of the figure magnified for clarity. 
\label{fig:prospects}}
\end{figure}

SNIa discovered over the most recent 40 years (1970-2009) are shown in Figure~\ref{fig:catalog} as a function of distance. We see that while rare, there have nonetheless been a number of SNIa within 10 Mpc, at approximately 1 per decade (unfortunately, none in the recent decade). It is clear that one needs to go only a small factor in distance to observe $\gtrsim 10$ SNIa per decade. At larger distance, we see that the commencement of dedicated SNIa searches in the 1990s dramatically increased the number of SNIa discovered. Yet these are still underestimates, due to missing coverage at large distances. Furthermore the $4\pi$ sky is not evenly sampled; the Northern hemisphere is more closely observed than the South, resulting in a SNIa discovery ratio of approximately $1.4:1$ (2000-2009, within 100 Mpc). 

The comparison to the cosmic SNIa rate is also revealing. The red lines show the extrapolated cosmic SNIa rate, and the incompleteness of the catalog is apparent by $\sim 30$~Mpc. With next-generation surveys such as the Palomar Transient Factory \citep{2009PASP..121.1395L}, SNIa measurements are becoming more complete. Many more SNIa are also being discovered pre-maximum, offering more targets for future gamma-ray detectors. 

Around 20~Mpc the catalog shows more SNIa than the cosmic extrapolation. The excess is as high as a factor $\sim 3$ for the 20--22 Mpc bin, although this is not statistically strong. Even considering the uncertainty in the cosmic SNIa rate this excess persists. There is also some excess from the Virgo galaxy cluster, located at $\sim 18$ Mpc in the Northern hemisphere and containing in excess of 1000 galaxies. The SNIa sample of \cite{2009MNRAS.395.1409S}, which contains SNIa within 28~Mpc, similarly shows a high SNIa rate.

\subsection{Review of SNIa gamma-ray observations \label{sec:localpast}}

There are strong upper limits on the gamma-ray emission from individual SNIa. Among the earliest was SN~1986G, observed by the Solar Maximum Mission (\emph{SMM}) instrument  \citep{1990ApJ...362..235M}, followed by SN~1991T \citep{1994A&A...292..569L,1995ApJ...450..805L,1997AIPC..410.1084M} and SN~1998bu \citep{1999HEAD....4.0803L} both observed by the Compton Gamma Ray Observatory (\emph{CGRO}). The derived limits depend on the fact that SN~1986G was subluminous, SN~1991T was superluminous, and SN~1998bu was normal. Furthermore, distance uncertainties weaken the limits. In all cases, limits are compatible with theoretical predictions within uncertainties, though very close \citep{1994ApJS...92..501H}. 

On the top x-axis of Figure~\ref{fig:catalog}, the distance scale is converted to a 847~keV number flux, assuming that all SNIa yield $S_\gamma = 4.3 \times 10^{47} $~s$^{-1}$ during the peak $10^6$ s. This is a conservative estimate for a normal SNIa; subluminous and superluminous SNIa would have different axis scalings (Table~\ref{table:gamma}). The in-flight 3$\sigma$ narrow-line sensitivity of \emph{SPI} onboard the \emph{INTEGRAL} satellite is $3 \times 10^{-5}  $ cm$^{-2}$~s$^{-1}$ at $\sim 1$ MeV for a $10^6$~s exposure \citep{2003A&A...411L..91R}. The equivalent for \emph{COMPTEL} onboard the \emph{CGRO} satellite is $6 \times 10^{-5} $ cm$^{-2}$~s$^{-1}$ \citep{1993ApJS...86..657S}. The Nuclear Compton Telescope (\emph{NCT}) balloon experiment has a narrow-line sensitivity comparable to \emph{INTEGRAL} \citep{2009ITNS...56.1250B}, the DUAL gamma-ray mission is 30 times more sensitive \citep{2010arXiv1006.2102B}, and the proposed Advanced Compton Telescope (\emph{ACT}) satellite is 60 times better \citep{2006NewAR..50..604B}. 

Nearby SNIa are labeled in Figure~\ref{fig:catalog}, from which we see that those observed by \emph{CGRO} were just out of range. Note that SN~1991T was a superluminous SNIa, for which the \emph{CGRO} horizon is larger than shown in Figure~\ref{fig:catalog}, but still insufficient. SN~2003gs was briefly observed by the \emph{INTEGRAL} satellite, with no line detection \citep{2009arXiv0903.0772L}. 

The \emph{INTEGRAL} sensitivity horizon to 847 keV gamma rays is $\sim 11$~Mpc, corresponding to 0.16 SNIa per year (Figure \ref{fig:prospects}). The 812~keV flux is smaller, and is visible out to a maximum of 4--6~Mpc, corresponding to 0.006--0.03 SNIa per year. The 847 keV horizon for superluminous and subluminous SNIa are $\sim$13 Mpc and $\sim$7 Mpc, respectively. In practice, the energy resolution of \emph{SPI} is better than the Doppler broadened width of the lines, so that the signal is spread over multiple energy bands. This reduces the effectiveness of detector background rejection with \emph{SPI}, and the sensitivity horizon decreases by a factor of at most $\sim 2$ \citep{1998MNRAS.295....1G}. This issue similarly applies to next-generation detectors which we discuss in the next section. The rates are summarized in Table~\ref{table:number}, neglecting the reduced detector background rejection due to line width. These are also labeled in Figure \ref{fig:prospects}, where the cumulative number of catalog SNIa are shown as functions of the 847 keV flux (bottom x-axis) and distance (top x-axis)

\begin{table}[b]
\begin{center}
\caption{Per-year SNIa gamma-ray detection rates \label{table:number}}
\begin{tabular}{lccc}
\tableline\tableline
Search Mode & \multicolumn{3}{c}{Sensitivity [${\rm cm^{-2} \, s^{-1}}$] } \\ 
 & $3 \times 10^{-5}$ & $1 \times 10^{-5}$ & $5 \times 10^{-7}$ \\
\tableline
847 keV detection [yr$^{-1}$] 	& 0.1--0.2 			&  1.0--1.4 	& 60--100  \\
847 keV light curve [yr$^{-1}$]  	& 0.1--0.2  		&  1.0--1.3 	& 60--80  \\
812 keV detection [yr$^{-1}$]  	&  0.006--0.03 		&  0.03--0.2 	& 2--20 \\
812 keV light curve [yr$^{-1}$]  	&  0.001--0.006		&  0.005--0.03  	& 0.5--4  \\
\tableline
\end{tabular}
\tablecomments{Rates are derived using the cosmic SNIa rate and rates from the SAI catalog, whichever is larger. The sensitivities $3 \times 10^{-5} \, {\rm cm^{-2} \, s^{-1}}$ and $5 \times 10^{-7} \, {\rm cm^{-2} \, s^{-1}}$ correspond to the $3 \sigma$ sensitivity of \emph{INTEGRAL-SPI} \citep{2003A&A...411L..91R} and the baseline $3 \sigma$ sensitivity of \emph{ACT} \citep{2006NewAR..50..604B}, respectively. The intermediate sensitivity of $1 \times 10^{-5} \, {\rm cm^{-2} \, s^{-1}}$ is a hypothetical satellite for illustrative purposes. The terms ``detection'' and ``light curve'' are defined as a single $10^6$ s exposure detection and three independent $10^6$ s exposure detections, respectively. The difference between detection and light curve prospects reflects the fact that the 847 keV is flatter and 812 keV is more peaked at their respective peak fluxes.}
\end{center}
\end{table}

\subsection{Prospects for future SNIa gamma-ray detection \label{sec:local}}

The initial goal of SNIa gamma-ray studies would be a single long-exposure detection of the 847 keV line from a nearby SNIa, which, together with the optical data, would provide a robust handle on the amount of $^{56}$Ni synthesized in the explosion. The prospects for detecting the 847 keV emission from $\sim 1$ SNIa per year is not too far from current \emph{INTEGRAL} sensitivity (Table~\ref{table:number}). 

\begin{figure}[t]
\centering\includegraphics[width=\linewidth,clip=true]{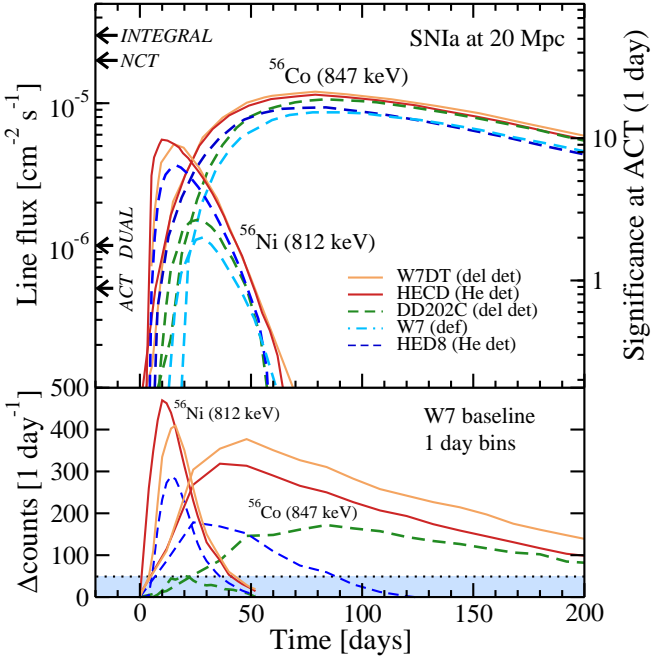}
\caption{Top: 812 keV and 847 keV light curves, for a selection of SNIa simulations (see Table \ref{table:gamma}) at a distance of 20 Mpc. The $3\sigma$ detector sensitivities ($10^6$ s) are labeled by arrows. On the right axis the statistical significance for \emph{ACT}, over a 1 day viewing period, is shown (see text for how significance accumulates with the square-root of the viewing period). \emph{ACT} will produce detailed 812 keV and 847 keV light curves of all SNIa at 20 Mpc (subluminous SNIa have not been plotted for clarity, but remain above the \emph{ACT} sensitivity at 20 Mpc). Bottom: the \emph{difference} between models, taking \emph{W7} as the reference, shown as the difference in signal count rate per day by \emph{ACT}. The horizontal shading indicates the square root of the background rate.
\label{fig:lightcurve}}	
\end{figure}

However, more important in the future is improving beyond a single 847 keV detection and measuring the gamma-ray light curve, as this ultimately holds the power to distinguish between SNIa models. In particular, the period during which the system transitions from an optically thick to a thin one gives the $p_{\rm esc}(t)$ in Eq.~\ref{eq:emission} and provides the best opportunity for model discrimination. For example, He-detonation produces $^{56}$Ni nearer to the WD surface compared to other models, resulting in an earlier transition and hence an earlier rise of gamma-ray lines. Deflagration is the other extreme, with ignition occurring in the central regions of the WD, resulting in slower rise of the gamma-ray emission. These points are clear in Figure~\ref{fig:lightcurve}, where the He-detonation models \emph{HECD} and \emph{HED8} are shown together with delayed-detonation (\emph{W7DT} and \emph{DD202C}) and deflagration (\emph{W7}) models. In addition, detection of the 812 keV line from $^{56}$Ni is important, given its strong emission during the transition period. In Figure~\ref{fig:lightcurve}, both super-luminous SNIa (solid) and normal SNIa (dashed) are shown together, but optical observations would distinguish between these two classes so we focus on model testing within each class. 

First, we show on the right y-axis of Figure~\ref{fig:lightcurve} the significance that can be achieved by \emph{ACT} observing a SNIa at 20 Mpc in 1-day ($\sim 10^5$ s) viewing periods. Since MeV gamma-ray satellites are dominated by detector backgrounds, the significance of the SNIa signal scales as $s = N_{\rm sig} / {N_{\rm bkg}}^{1/2} \propto T^{1/2} $, where $N_{\rm sig}$ is the signal counts, $N_{\rm bkg}$ is the background counts, and $T$ is the viewing period. The targeted $3 \sigma$ sensitivity of \emph{ACT} is $5 \times 10^{-7} \, {\rm cm^{-2} \, s^{-1}}$ (for $10^6$ s), and we scale it to a viewing period of 1 day. We see that \emph{ACT} will allow detailed reconstruction of the 812 keV and 847 keV light curves with very large significance at this distance. 

Second, provided that the SNIa is close enough for sufficient photon statistics, studying the shape of the light curve does not depend on knowing the exact distance to the SNIa. The shape is distinctly different between models, as we illustrate more clearly in the bottom panel of Figure~\ref{fig:lightcurve}, where we show the difference counts at \emph{ACT}, for a viewing period of 1 day, all relative to the $W7$ model. An uncertain distance would scale the fluxes accordingly, but retain the shape differences between models. For illustration, the square root of the background counts is shown as the horizontal shading. \emph{ACT} would easily distinguish between normal SNIa models with high significance at this distance. The difference between superluminous SNIa is smaller than the square root of the background counts; better testing would be possible with a coarser time binning. 

The physics potential can be scaled to a detector with generic properties. The significance shown in Figure~\ref{fig:lightcurve} (top right y-axis) scales as
\begin{eqnarray}
s &\approx& 17.6 \, \left( \frac{ \phi_{\rm sig}}{  1 \times 10^{-5} \, {\rm cm^{-2} \, s^{-1}} } \right)  \left( \frac{ T }{ 1 \, {\rm day} } \right)^{1/2} \\ \nonumber
&& \times \left( \frac{5 \times 10^{-7} \, {\rm cm^{-2} \, s^{-1}} }{\phi_{\rm sens}} \right)  \left( \frac{10^6 \, {\rm s} }{ T_0 } \right)^{1/2},
\end{eqnarray}
where $\phi_{\rm sig}$ is the signal flux, $T$ is the viewing period, and $\phi_{\rm sens}$ is the $3 \sigma$ line sensitivity in a $T_0=10^6$~s viewing period. Note significance accumulates with the square-root of the time period, i.e., 2 consecutive $10 \sigma$ detection implies an overall $10 \sqrt{2} \sigma$ detection. The square root of the background counts (bottom of Figure \ref{fig:lightcurve}) scales as
\begin{eqnarray}
\sqrt{N_{\rm bkg}} &\approx& 50 \, \left( \frac{ T }{ 1 \, {\rm day} } \right)^{1/2}  \left( \frac{A_{\rm eff}}{10^3 \, {\rm cm^{2}}} \right)  \\ \nonumber
&& \times \left( \frac{ \phi_{\rm sens}}{5 \times 10^{-7} \, {\rm cm^{-2} \, s^{-1}}} \right) 
\left( \frac{T_0}{10^6 \, {\rm s} } \right)^{1/2}
\end{eqnarray}
where $A_{\rm eff}$ is the effective area of the detector. 

The rate of SNIa within 20~Mpc is $\sim 1$ per year, and therefore, \emph{ACT} will strongly constrain SNIa models at a rate of at least 1 per year. SNIa further away also yield information, with statistical significance gained by coarser time binning. At 2--20 times per year, \emph{ACT} would detect the 812 keV emission and build a detailed 847 keV light curve. At 60-80 times per year, \emph{ACT} would detect the 847 keV light curve near peak, and at almost 100 times per year, it would detect the 847 keV emission and measure the variation in the $^{56}$Ni yield of SNIa. We summarize these gamma-ray detection prospects in Table~\ref{table:number}. The wide-field all-sky nature of \emph{ACT} could in principle detect SNIa independently of optical discoveries, although in practice next-generation optical surveys would bring out the full potential of \emph{ACT} by providing more targets for joint analysis. \emph{DUAL}'s focusing capabilities would rely on knowledge of the SNIa location from an optical trigger. 

For all the above estimates we have assumed the narrow-line sensitivities of detectors as documented. Since SNIa gamma-ray lines are expected to be Doppler-broadened by up to 3\%, this treatment is optimistic, although it does allow us to compare the maximum performance of detectors. In practice, the broad-line sensitivities are only marginally worse than the narrow-line sensitivities. For example, the $3$\% broadened line sensitivity of \emph{ACT} is $1.2 \times 10^{-6} \, {\rm cm^{-2} \, s^{-1}}$ ($3\sigma$ in $10^6$ s; \cite{2006NewAR..50..604B}), a factor 2 different from the narrow-line sensitivity. The gamma-ray horizon therefore decrease by $\sim 1/\sqrt{2}$, and the numbers in the final column of Table \ref{table:number} decrease to 20--30, 16-21, 1--5, and 0.1--1, respectively. Although smaller, these rates would still allow rapid progress in the understanding of SNIa. Similarly, the significance and $\sqrt{N_{\rm bkg}}$ for the broad-line case can be calculated to be $\approx 7.3$ and $\approx 120$ in a 1-day bin, respectively. The significance remains high, and $\sqrt{N_{\rm bkg}}$ remains smaller than the model differences, demonstrating the physics potential.

\subsection{Discussions \label{sec:discussion}}

We focused on the 812 keV and 847 keV lines because they are expected to have the highest fluxes. There are other lines for which there are interesting prospects, for example the 158 keV line from $^{56}$Ni and the 1238 keV line from $^{56}$Co. The detectability of these lines is only marginally less than that of the 812 keV and 847 keV lines (Eqs.~\ref{eq:Ni}--\ref{eq:Co}). Additionally, the 511 keV line from annihilation of positrons produced in $^{56}$Co decay (Eq.~\ref{eq:Co}) is particularly important for understanding the positron escape fraction, thought to be $\sim 1$\%, but quite uncertain \citep{2001ApJ...559.1019M,2006AJ....132.2024L}. It has important implications for the origin of the Galactic 511 keV emission \citep[see, e.g.,][and references therein]{2006PhRvL..97g1102B}. It should also be mentioned that other radioactive nuclei such as $^{44}$Ti and $^{57}$Ni are expected in SNIa. The half life of $^{44}{\rm Ti} \to \, ^{44}{\rm Sc}$ is 68 years, implying historical SNIa as prime targets rather than the SNIa discussed in this paper. The $^{57}{\rm Ni} \to \, ^{57}{\rm Co} \to \, ^{57}{\rm Fe}$ decay chain, with half lives of 52 hours and 391 days, would be a more suitable target. However, as the production of $^{57}$Ni is not as abundant as $^{56}$Ni in most cases, it would be a useful line only for exceptionally close SNIa. 

Other model-distinguishing features that we have not mentioned include line shift and line width evolutions. During the first few weeks, the high optical depth means that much of the gamma-ray emission originates from the approaching ejecta, with photons escaping essentially radially, so that the lines are blue-shifted. Together with the line width, these are measures of the expansion velocity, and their time evolution tests SNIa models. The caveat with the former is that shifts occur mostly during the optically thick regime, so model testing is difficult except for exceptional cases (e.g., the DD scenario we mention next). The line width offers a more promising test, in particular with detectors with spectral resolution of order $\lambda / \delta \lambda \sim 200$--300 \citep[see, e.g.,][for detailed predictions]{1998ApJ...492..228H,2004ApJ...613.1101M}.

The gamma-ray emission for the DD scenario experiences significant suppression due to the envelope resulting from the merging WDs. In the \emph{det2env4} model of \cite{1998ApJ...492..228H}, the enveloping matter means that the line shifts remain high for a long period of time, up to $\sim 100$ days, whereas SD scenario models all fall in a few weeks \citep{1998ApJ...492..228H}. In addition, the envelope strongly suppresses the 812 keV from $^{56}$Ni decay, so that the flux is weaker than even the deflagration model \emph{W7}. The 847 keV emission also peaks at a later stage. Recently, several authors have explored the collision of two WDs in dense stellar systems as an alternate pathway to SNIa within the family of DD scenarios \citep{2009ApJ...705L.128R,2009MNRAS.399L.156R}. Shock-triggered thermonuclear explosions in such events are predicted to yield $^{56}$Ni masses that are sufficient to power subluminous SNIa. Although detailed gamma-ray light curves have not yet been published, they could be targets for next-generation gamma-ray detectors if their gamma-ray emissions are comparable to those of subluminous SNIa.

\subsubsection{SNIa Variants}

Growth of the SNIa sample has led to variants of SNIa being discovered. Deep searches have revealed peculiar faint explosions: SN~2002bj, observed by the LOSS survey, was a faint SNIa showing unusually rapid evolution on time scales of days \citep{2010Sci...327...58P}. The low luminosity and short rise time ($<$ 7 days) translate to 0.15--0.25${\rm M_\odot}$ of $^{56}$Ni. Another supernova, SN~2005E, has an estimated rise time of 7--9 days, very small estimated $^{56}$Ni yield, and spectra showing abundance of helium burning products \citep{2010Sci...327...58P}. Its location in an isolated galaxy with little star formation activity suggests an old progenitor, e.g., WD binaries, instead of the core collapse of a massive star. These resemble so-called ``.Ia'' explosions, which has been proposed to occur in binary WDs undergoing helium mass transfer \citep{2007ApJ...662L..95B}. As the binary evolves, the mass required for unstable helium burning increases, until a final flash that leads to a faint thermonuclear explosion that is one-tenth as bright for one-tenth the time of a normal SNIa \citep{2009ApJ...699.1365S,2009ApJ...707..193F,2010ApJ...715..767S}. The predicted rates match those observed, both being in the range of a few percent of SNIa. 

Although the $^{56}$Ni yield is lower than in a normal SNIa, this is compensated by a larger escape probability for gamma rays. Indeed, the rapidly falling light curves of SNIa variants suggest that of order unity of the $^{56}$Co decay gamma rays are escaping already by $\sim 20$ days. This contrasts with normal SNIa, where up to 90\% of the gamma rays at 20 days are down-scattered in the ejecta. It turns out the low $^{56}$Ni yield and high escape fraction almost compensate each other. The time-integrated escape fraction of gamma-ray photons in normal SNIa is about 50\%, so that SNIa variants, with half as much $^{56}$Ni and up to twice the total escape fraction, provide a comparable gamma-ray output to normal SNIa. The low rate of SNIa variants therefore make them subdominant contributors to gamma-ray observations. 

The situation is potentially more interesting for positrons. Adopting the canonical escape fraction 1\% for normal SNIa \citep{1999ApJS..124..503M}, SNIa variants could contribute up to $\sim 2$ times more than normal SNIa towards the positron budget, depending on their rate. This would help narrow the gap between the supernovae positron yield and that required from observations of the Galactic 511 keV line. Again, line detection will prove important: late-time detection of the 511 keV line would provide a direct constraint of the fraction of positrons escaping through the ejecta. 

Another SNIa variant is the extremely luminous super-Chandrasekhar mass SNIa. They exhibit higher magnitudes and slower ejecta velocities compared to normal SNIa, with predicted progenitor masses that exceed the Chandrasekhar mass. For example, SN~2003fg was a $M_V=-19.9$ SNIa with a predicted Ni yield of $\sim 1.3 \, {M_\odot}$ and progenitor mass of  $\sim 2 \, {M_\odot}$  \citep{2006Natur.443..308H}. The recent super-Chandrasekhar mass SN~2009dc has a Ni yield of $1.6 \, {\rm M_\odot}$ \citep{2009ApJ...707L.118Y}, and spectropolarimetry observations suggest the explosion was near spherical, supporting a truly super-Chandrasekhar progenitor \citep{2010ApJ...714.1209T}. Rapid rotation may support such a massive WD. Binary WDs could also produce super-Chandrasekhar mass SNIa. If the light curves of candidate super-Chandrasekhar mass supernovae are indeed powered by a larger-than-normal amount of produced nickel, this will give a large gamma-ray signal; if they are instead powered by circumstellar interactions, this will not give such a large gamma-ray signal. 

\section{Conclusions}\label{sec:summary}

Gamma rays provide unique clues to the currently debated progenitor properties and explosion mechanisms of Type Ia supernovae (SNIa). They directly probe the power source of SNIa and provide tomography of the SNIa ejecta. As the importance of SNIa in astrophysics and cosmology continues to grow, the detection of gamma rays become increasingly essential. The detectability relies on there being sufficient optical SNIa discoveries within the ``sensitivity horizon'' of gamma-ray telescopes. In this paper we investigate the prospects for studying SNIa physics using current and future gamma-ray detectors. Below we summarize our results.

\subsection{Results on SNIa rates and SNIa progenitors}

We first investigate the SNIa rate, which is a prerequisite for SNIa gamma ray detection prospects. It is also interesting since the SNIa rate with respect to its progenitor formation rate depends on what the SNIa progenitors are. We jointly analyze the cosmic star formation rate, cosmic SNIa rate, and the SNIa rate derived from SNIa catalogs. We deduce a delay time distribution (DTD) and SNIa fraction that fit the data, and discuss the local ($< 100$ Mpc volume) SNIa rate.

\begin{itemize}

\item \emph{Delay-times}: When SNIa rate measurements are categorized according to sample size and fraction of spectroscopically-identified SNIa, we find that the more reliable measurements collectively show significantly slower evolution with redshift than the star formation rate. The difference is due to the delay between progenitor formation and SNIa: we find that a DTD of the form $\propto t^{-\alpha}$ with $\alpha=1.0 \pm 0.3$ provides a good fit. The substantially-prompt bimodal DTD of \cite{2006MNRAS.370..773M} and the narrow Gaussian DTD around 3.4 Gyr of \cite{2008ApJ...681..462D} do not fit the global data as well.

\item \emph{SNIa efficiency}: For our DTD, we find an efficiency of making SNIa of $(5\pm1) \times 10^{-4} \, { \rm M_\odot^{-1}}$. Assuming that the SNIa progenitor mass range is 3--8 ${\rm M_\odot}$, this equates to a SNIa fraction of $2.4\pm0.5$\% (for the Salpeter initial mass function; the dependence on this choice is weak).

\item \emph{Implication for local SNIa rate}: The local SNIa rate is much higher than previously thought. Supernova catalog entries between 2000 to 2009 reveal on average 40 SNIa per year within 100 Mpc, a significant increase from the $\sim 2$ in the 1980s \citep{1987ApJ...322..215G} and 5.5 in the 1990s \citep{1997ApJ...489..160T}. Even so, discoveries are still severely incomplete outside about 30 Mpc. The expected true rate within 100 Mpc is about $\approx 100$ SNIa per year. The SNIa rate within $20$ Mpc is $\sim 1$ per year.

\end{itemize}

\subsection{Results on SNIa gamma rays and SNIa explosions}

The detection of gamma rays from SNIa directly tests the $^{56}$Ni mass inferred from optical observations, and also provides tomography of the SNIa interior by measuring the time-dependent escape probability. In the previous section we quantitatively discussed how SNIa discoveries are becoming more complete, and highlighted 20 Mpc as the distance for annual SNIa discovery. Our main results on SNIa gamma-ray detection prospects and the physics potential of gamma-ray detectors are as follows. 

\begin{itemize}

\item \emph{CGB contribution}: The SNIa contribution to the cosmic gamma-ray background (CGB) is at most 10--20\% of the CGB flux published by \emph{SMM} and \emph{COMPTEL}, confirming previous results. The origin of the $\sim$ MeV CGB, and whether the SNIa contribution can be identified, therefore remain an important task to be clarified by future gamma-ray observations. 

\item \emph{Current local SNIa prospects}: Local SNIa gamma-ray detection prospects are better than thought a decade ago, principally driven by the vastly increased rate of SNIa discoveries. Current gamma-ray satellites probe SNIa in the rare regime: \emph{INTEGRAL} probes SNIa within $\sim 10$~Mpc, occurring at a rate of $\approx 0.1$ SNIa per year. This is only somewhat larger than previous estimates of $\sim 0.03$ per year \citep{1987ApJ...322..215G,1997ApJ...489..160T}, but we are more confident about the normalization. 

\item \emph{Future local SNIa prospects}: The distance for annual SNIa discovery (20 Mpc) is only a factor 2 further than current horizons (Figures~\ref{fig:catalog} and \ref{fig:prospects}). Future detectors with a line sensitivity of $1 \times 10^{-5} \, {\rm cm^{-2} \, s^{-1}}$---only a factor 3 better than that of \emph{INTEGRAL}---will cross this threshold. The proposed \emph{ACT} satellite, with a 60 times better line sensitivity than that of \emph{INTEGRAL}, will probe SNIa out to $\sim 90$ Mpc, translating to a rate of 100 SNIa per year. Improved supernova surveys are discovering more SNIa and at earlier times, making this possible. 

\item \emph{Implication for explosions}: Nearby SNIa will be targets for detecting gamma ray lines from both $^{56}$Ni and $^{56}$Co decays, and for their light curve reconstruction. The \emph{ACT} satellite will give a hugely significant $> 100 \sigma$ detection of gamma-ray light curves every year from SNIa within 20 Mpc. These SNIa will allow detailed analysis of SNIa explosion physics, providing unprecedented understanding of the SNIa explosion mechanism. A more modest detector with a line sensitivity of $1 \times 10^{-5} \, {\rm cm^{-2} \, s^{-1}}$ would measure the 847 keV light curve for tomography once a year, and detect the 812 keV line $\sim 0.1$ per year (Table \ref{table:number}). 
\end{itemize}

Given the importance of SNIa from cosmology to nucleosynthesis, the need to understand the mechanisms of SNIa will only increase with time. Detecting their gamma ray emission is the only way to probe their inner physics. Judging from the SNIa rates, even a modest improvement in gamma-ray line sensitivity would reach the distance range with annual detections. Proposed next-generation gamma-ray satellites are in an excellent position to rapidly offer revolutionary results.


\acknowledgments
We are grateful to Louie Strigari for collaboration on an early stage of this project, and to Steven Boggs, Tomas Dahlen, Alex Filippenko, Neil Gehrels, Laura Greggio, Matt Kistler, Dan Maoz, Ken'ichi Nomoto, Evan Scannapieco, Stephen Smartt, Kris Stanek, Louie Strigari, Mark Sullivan, Frank Timmes, and Haojing Yan for sharing their expertise and advice. SH and JFB were supported by NSF CAREER Grant PHY-0547102 (to JFB).



\clearpage

\begin{deluxetable*}{lcccrl}
\tablecaption{Type Ia supernova rate measurements \label{table:data}} 
\tablehead{ 
	&	\multicolumn{2}{c}{SNIa rate (with stat.~and syst.~errors)}	&	 \multicolumn{3}{c}{Survey information} 	\\
  z 	&   [$h_{70}^2$ SNu]   &  [$10^{-4}$ $h_{70}^3$ yr$^{-1}$ Mpc$^{-3}$]   &   $N_{Ia}$   &   \% spec-ID   & Ref. 
} 
\startdata
\hline
(28 Mpc)  &  -  &  $0.38\pm0.06$  &  37  &  100  &  \cite{2009MNRAS.395.1409S} \\
(40 Mpc)  &  $0.18\pm0.05$  &  -  &  70  &  100  &  Cappellaro et al.~(1999) \\

0  &  -  &  $0.301^{+0.038+0.049}_{-0.037-0.049}$  &  274  &  -  &  \cite{2010arXiv1006.4613L}  \\
0.09  &  - &  $0.293^{+0.090+0.017}_{-0.071-0.004}$  &  17  &  100  &  \cite{2008ApJ...682..262D} \\
0.1  &  $0.20\pm0.1$  &  -  &  19  &  100  &  \cite{2003ApJ...599L..33M} \\
0.13  & $0.125^{+0.044+0.028}_{-0.034-0.028}$ &  $0.199^{+0.070+0.047}_{-0.054-0.047}$  &  14  &  100  & Blanc et al.~(2004) \\
0.14  & $0.22^{+0.17+0.06}_{-0.10-0.03}$ &  $0.34^{+0.27+0.11}_{-0.16-0.06}$  &  4  &  100  & Hardin et al.~(2000)\tablenotemark{a} \\
0.46  & -   &  $0.48\pm0.17$  &  8  &  100  & \cite{2003ApJ...594....1T} \\
0.47  & $0.154^{+0.039+0.048}_{-0.031-0.033}$   &  $0.42^{+0.06+0.13}_{-0.06-0.09}$  &  73  &  100  &  \cite{2006AJ....132.1126N}  \\
0.55  & $0.28^{+0.05+0.05}_{-0.04-0.04}$   &  $0.53^{+0.10+0.11}_{-0.09-0.11}$  &  38  &  100  &  \cite{2002ApJ...577..120P}   \\

{[0.2, 0.6]} &  -  &  $0.80^{+0.37+1.66}_{-0.27-0.26}$  &    8  &  54  &  \cite{2008ApJ...681..462D}   \\
{[0.6, 1.0]} &  -  &  $1.30^{+0.33+0.73}_{-0.27-0.51}$  &  25  &  54  &  \cite{2008ApJ...681..462D}   \\
{[1.0, 1.4]} &  -  &  $1.32^{+0.36+0.38}_{-0.29-0.32}$  &  20  &  54  &  \cite{2008ApJ...681..462D}   \\
{[1.4, 1.8]} &  -  &  $0.42^{+0.39+0.19}_{-0.23-0.14}$  &    3  &  54  &  \cite{2008ApJ...681..462D}   \\

{[0.2, 0.6]}  &  -  &  $0.69^{+0.34+1.54}_{-0.27-0.25}$  &    3  &  52  &  \cite{2004ApJ...613..189D}   \\
{[0.6, 1.0]}  &  -  &  $1.57^{+0.44+0.75}_{-0.25-0.53}$  &  14  &  52  &  \cite{2004ApJ...613..189D}   \\
{[1.0, 1.4]}  &  -  &  $1.15^{+0.47+0.32}_{-0.26-0.44}$  &    6  &  52  &  \cite{2004ApJ...613..189D}    \\
{[1.4, 1.8]}  &  -  &  $0.44^{+0.32+0.14}_{-0.25-0.11}$  &    2  &  52  &  \cite{2004ApJ...613..189D}   \\

{[0.025, 0.050]}  &  -  &  $0.278^{+0.112+0.015}_{-0.083-0.00}$  &  516\tablenotemark{b}  &  52  &  \cite{2010ApJ...713.1026D} \\
{[0.075, 0.125]}  &  -  &  $0.259^{+0.052+0.018}_{-0.044-0.001}$  &  516\tablenotemark{b}  &  52  &  \cite{2010ApJ...713.1026D}  \\
{[0.125, 0.175]}  &  -  &  $0.307^{+0.038+0.035}_{-0.034-0.005}$  &  516\tablenotemark{b}  &  52  &  \cite{2010ApJ...713.1026D}  \\
{[0.175, 0.225]}  &  -  &  $0.348^{+0.032+0.082}_{-0.030-0.007}$  &  516\tablenotemark{b}  &  52  &  \cite{2010ApJ...713.1026D}  \\
{[0.225, 0.275]}  &  -  &  $0.365^{+0.031+0.182}_{-0.028-0.012}$  &  516\tablenotemark{b}  &  52  &  \cite{2010ApJ...713.1026D}  \\
{[0.275, 0.325]}  &  -  &  $0.434^{+0.037+0.396}_{-0.034-0.016}$  &  516\tablenotemark{b}  &  52  &  \cite{2010ApJ...713.1026D}  \\

0.3  &  $0.22^{+0.10+0.16}_{-0.08-0.14}$  &  $0.34^{+0.16+0.21}_{-0.15-0.22}$  &  26  &  35  &  Botticella et al.~(2008)\\

0.25  &  -  &  $0.17\pm0.17$  &    1  &  24  &  \cite{2006ApJ...637..427B} \\
0.35  &  -  &  $0.53\pm0.24$  &    5  &  24  &  \cite{2006ApJ...637..427B} \\
0.45  &  -  &  $0.73\pm0.24$  &    9  &  24  &  \cite{2006ApJ...637..427B} \\
0.55  &  -  &  $2.04\pm0.38$  &  29  &  24  &  \cite{2006ApJ...637..427B} \\
0.65  &  -  &  $1.49\pm0.31$  &  23  &  24  &  \cite{2006ApJ...637..427B} \\
0.75  &  -  &  $1.78\pm0.34$  &  28  &  24  &  \cite{2006ApJ...637..427B} \\

0.2  &  $0.14^{+0.03+0.03}_{-0.03-0.03}$  &  $0.189^{+0.042+0.046}_{-0.034-0.045}$  &  17  &  0  &  \cite{2008MNRAS.389.1871H} \\

{[0.0, 0.5]} &  -  &  $0.0^{+0.24}_{-0.00}$    &    0.0   &  0  & \cite{2007MNRAS.382.1169P} \\
{[0.5, 1.0]} &  -  &  $0.43^{+0.36}_{-0.32}$    &    5.5   &  0  & \cite{2007MNRAS.382.1169P} \\
{[1.0, 1.5]} &  -  &  $1.05^{+0.45}_{-0.56}$  &  10.0  &  0  & \cite{2007MNRAS.382.1169P} \\
{[1.5, 2.0]} &  -  &  $0.81^{+0.79}_{-0.60}$    &    3.0   &  0  & \cite{2007MNRAS.382.1169P} \\

{[0.2, 0.6]} &  -  &  $0.53^{+0.39}_{-0.17}$   &   5.44  &  0  & \cite{2008ApJ...673..981K} \\
{[0.6, 1.0]} &  -  &  $0.93^{+0.25}_{-0.25}$   &   18.33  &  0  & \cite{2008ApJ...673..981K} \\
{[1.0, 1.4]} &  -  &  $0.75^{+0.35}_{-0.30}$   &   8.87  &  0  & \cite{2008ApJ...673..981K} \\
{[1.4, 1.7]} &  -  &  $0.12^{+0.58}_{-0.12}$   &   0.35  &  0  & \cite{2008ApJ...673..981K} \\

\tableline
\tablenotetext{a}{Hardin et al. (2000) uses the EROS SN sample (1997), with $N_{Ia} = 4$. To derive the volumetric value we use their SNu value with the 2dF luminosity density, with the error on the luminosity density added in quadrature to the systematics.}
\tablenotetext{b}{Dilday et al. (2010) uses SDSS-II supernova survey data with a total $N_{Ia}=516$.}
\tablecomments{Redshifts or distance noted are representative, not exact. All values have been corrected to $h=0.7$ and to match our ($\Omega_M$, $\Omega_\Lambda$)=(0.3, 0.7) cosmology. We quote rates in units of SNu (SNu $= SN (100 \, {\rm yr})^{-1} (10^{10} \, {\rm L_\odot^B})^{-1} $) or yr$^{-1}$ Mpc$^{-3}$, whichever is reported by the authors. The quoted errors are statistical followed by systematic; where only one is present, the systematic is the one not available. $N_{\rm Ia}$ is the number of SNIa used in the study, and the percentage of $N_{\rm Ia}$ that are spectroscopically identified as SNIa are shown.}
\end{deluxetable*}

\end{document}